\documentclass{ckm}                 

\confname{Workshop on the CKM Unitarity Triangle, IPPP Durham, April
  2003}

\title{Worries and Hopes for SUSY in CKM Physics: The $b \to s$ Example}

\author{M Ciuchini\addressmark{a}, E Franco\addressmark{b}, 
 A Masiero\addressmark{c} and L
 Silvestrini\addressmark{b}}

\address[a]{INFN
    Sezione di Roma III and Dip. di Fisica, Univ. di Roma Tre, 
    Via della Vasca Navale 84, I-00146 Rome, Italy.}
\address[b]{INFN Sezione di Roma and Dip. di Fisica, Univ.
    di Roma ``La Sapienza'', P.le A. Moro 2,
    I-00185 Rome, Italy.}
\address[c]{Dip. di Fisica ``G.
    Galilei'', Univ. di Padova and INFN, Sezione di
    Padova, Via Marzolo 8, I-35121 Padua, Italy.}

\begin{document}

\begin{abstract}
We discuss the twofold r\^ole of flavor and CP violation as a
constraint in model building and as a signal of SUSY. Considering as
an example $b \to s$ transitions, we analyze present bounds on SUSY
parameters, discuss possible deviations from SM predictions in $B_d$
and $B_s$ physics and present strategies to reveal SUSY signals in
present and future experiments in the CKM domain.
\end{abstract}

\maketitle


\section{Introduction}
\label{sec:intro}

Right since the advent of SUSY in the phenomenological arena in the
early 80s , we have witnessed the presence of a twofold attitude of
SUSY searchers towards FCNC and/or CP violation: on one hand, FCNC
rare processes constitute a severe naturalness threat to low-energy
SUSY extensions of the SM (i.e., the so-called flavor and CP violation
problem of low-energy SUSY), whilst on the other hand they represent a
promising way to reveal the presence of new physics through the
effects of virtual SUSY particles running in the loops (and, in any
case, they constitute the most powerful tool we have at disposal to
constrain the enormous 124-parameter space of MSSM).

Concerning the former aspect, namely the FCNC threat, it is clear that
in these last 20 years we saw an increased success of the CKM flavor
pattern of the SM in reproducing the broad variety of flavor physics
data. The combined experimental and theoretical precisions that we
achieved have allowed for an enhanced confidence in the SM as the
correct explanation for all the observed rare processes so far. As for
possible SUSY contributions to observed FCNC and/or CP violation
phenomena, they have to represent a small perturbation of the main
bulk contribution represented by the CKM SM physics
\cite{hall}-\cite{Becirevic:2001jj} (alternatively, it is still
conceivable that we could have some very large new physics
contribution, but then a strong conspiracy in exactly reproducing the
SM expectations should be invoked \cite{gino}). The goodness of the SM
Unitarity Triangle fit naturally poses a challenging question for new
physics, namely should we ask for its complete flavor blindness to
avoid unnatural fine-tunings in coping with the flavor problem?

We think that it is premature to answer positively to the above
question. Indeed, no matter how good the UT fit looks so far within
the SM, we believe (and, indeed, well substantiate such faith
providing a few examples also in this talk) that there still exists a
relatively ample space for new physics departures from the SM
expectations in flavor physics. In other words, one should not forget
that as useful as the UT approach proves to be in encoding a large
amount of information concerning flavor physics, one should not
overestimate its power in constraining new physics effects. For
instance, a closer look at the UT fit reveals that new physics
contributions to $s \to d$ and $b \to d$ transitions are strongly
constrained, while new contributions to $b \to s$ transitions do not
affect the fit at all, unless they interfere destructively with the SM
amplitude for $B_s - \bar B_s$ mixing, bringing it below the present
lower bound of $14$ ps$^{-1}$ \cite{noickm}.

Hence, we consider strict low-energy flavor blindness a too strong
constraint. We think that it might be more advisable to insist on a
high-energy flavor blindness, namely flavor universality of the soft
breaking terms at the high energy scale where they show up in
supergravity theories. Such weak form of flavor blindness is by no
means equivalent to the strong requirement of low-energy flavor
universality. In the long running path from the superlarge scale at
which the SUSY soft breaking terms appear (presumably close to the
Planck scale) down to the electroweak scale, many factors can give
rise to even severe departures from flavor blindness of the low-energy
structure of the soft breaking terms. A clear example of such
difference between weak and strong flavor blindness arises in SUSY-GUTs
where the large mixing(s) in the neutrino sector may imply the
presence of large mixing angles in the right-handed down-type quarks
in spite of the initial flavor universality of the soft breaking terms
at scales above the GUT scale \cite{Moroi:2000tk,Chang:2002mq}.

As we mentioned above there is also the half-full glass perspective
when looking at the increased experimental and theoretical accuracy in
FCNC and CP violating processes and at its consequences for low-energy
SUSY. We are now in a situation allowing us to be optimistic on the
prospects to single out new-physics FCNC contributions in the present
and coming experiments on rare decays. If this is certainly true for
Lepton Flavor Violating (LFV) processes, (we all know well that
observing one muon radiatively decaying into an electron would be an
unquestionable signature of new physics), this is becoming true for
the more difficult hadronic sector where theoretical and experimental
intricacies often hinder the efforts to clearly disentangle new
physics effects from the SM background. In this contribution to the
CKM Workshop we will try to put all the above considerations into play
considering the particularly interesting sector of the $b \to s$
transitions.

\section{SUSY effects in \boldmath$b \to s$ transitions}
\label{sec:bs}

We now briefly report the results of the analysis of
ref.~\cite{Ciuchini:2002uv}, which aims at determining the allowed
regions in the SUSY parameter space governing $b \to s$ transitions,
studying the correlations among different observables and pointing out
possible signals of SUSY. The constraints on the parameter space come
from:
\begin{enumerate}
\item The BR$(B \to X_s \gamma)=(3.29 \pm 0.34)\times 10^{-4}$
  (experimental results as reported in~\cite{Stocchi:2002yi}, rescaled
  according to ref.~\cite{Gambino:2001ew}).
  
\item The CP asymmetry $A_{CP}(B \to X_s \gamma)=-0.02 \pm
  0.04$~\cite{Stocchi:2002yi}.
  
\item The BR$(B \to X_s \ell^+ \ell^-)=(6.1 \pm 1.4 \pm 1.3)\times
  10^{-6}$~\cite{Stocchi:2002yi}.
  
\item The lower bound on the $B_s - \bar B_s$ mass difference $\Delta
  M_{B_s} > 14.4$ ps$^{-1}$ \cite{Stocchi:2002yi}.
\end{enumerate}
We have also considered BR's and CP asymmetries for $B \to K \pi$ and
found that, given the large theoretical uncertainties, they give no
significant constraints on the $\delta$'s. For $B \to \phi K_s$, we
have studied the BR and the coefficients $C_{\phi K}$ and $S_{\phi K}$
of cosine and sine terms in the time-dependent CP asymmetry. In this
channel, the measured BR is somewhat larger than the SM prediction, which
would slightly favour SUSY contributions. However, given
the large errors, we prefer not to use it to constrain the SUSY
parameter space in our analysis.

All the details concerning the treatment of the different amplitudes
entering the analysis can be found in ref.~\cite{Ciuchini:2002uv}. In
summary, we use:

{\it i) $\Delta B=2$ amplitudes.} Full NLO SM and LO gluino-mediated
matching condition \cite{Gabbiani:1996hi}, NLO QCD evolution
\cite{Ciuchini,Buras:2001ra} and hadronic matrix elements from lattice
calculations \cite{Becirevic:2001yv}. See ref.~\cite{Gabrielli:2002fr}
for a discussion of the impact of chargino-mediated contributions in
$\Delta B=2$ processes.

{\it ii) $\Delta B=1$ amplitudes.} Full NLO SM and LO gluino-mediated
matching condition \cite{match,Cho:1996we} and NLO QCD evolution
\cite{NLODB1,Chetyrkin:1996vx,Misiak:bc}. The matrix elements of
semileptonic and radiative decays include $\alpha_s$ terms, Sudakov
resummation, and the first corrections suppressed by powers of the
heavy quark masses \cite{bsgme}. For non-leptonic decays, such as $B
\to K \pi$ and $B \to \phi K_s$, we adopt BBNS
factorization~\cite{BBNS}, with an enlarged range for the annihilation
parameter $\rho_A$, in the spirit of the criticism of
ref.~\cite{Charming}. This choice maximizes the sensitivity of the
factorized amplitudes to SUSY contributions, which is expected to be
much lower if the power corrections are dominated by the ``charming
penguin'' contributions~\cite{Ciuchini:1997hb}.  Another source of
potentially large SUSY effects in $B \to \phi K_s$ is the contribution
of the chromomagnetic operator which can be substantially enhanced by
SUSY without spoiling the experimental constraints from $B\to
X_s\gamma$~\cite{c8g}.  Indeed, the time-dependent asymmetry in $B \to
\phi K_s$ is more sensitive to the SUSY parameters in the case of
chirality-flipping insertions which enter the amplitude in the
coefficient of the chromomagnetic operator.  One should keep in mind,
however, that the corresponding matrix element, being of order
$\alpha_s$, has large uncertainties in QCD factorization.

We performed a MonteCarlo analysis, generating weighted random
configurations of input parameters (see ref.~\cite{Ciuchini:2000de}
for details of this procedure) and computing for each configuration
the processes listed above. We study the clustering induced by the
contraints on various observables and parameters, assuming that each
unconstrained $\delta_{23}^d$ fills uniformly a square $(-1\dots 1$,
$-1\dots 1)$ in the complex plane. The ranges of CKM parameters have
been taken from the UT fit ($\bar \rho=0.178 \pm 0.046$, $\bar
\eta=0.341 \pm 0.028$) \cite{ckm2}, and hadronic parameter ranges are
those used in ref.~\cite{Ciuchini:2002uv}.  Concerning SUSY
parameters, we fix $m_{\tilde q}=m_{\tilde g}=350$ GeV and consider
different possibilities for the mass insertions. In addition to
studying single insertions, we also examine the effects of the
left-right symmetric case $(\delta^d_{23})_{LL}=(\delta^d_{23})_{RR}$.

We stress that, having fixed the relevant SUSY masses and SM
parameters, the analysis we perform varying a single $\delta_{23}^d$
and computing various observables is completely analogous to the
standard UT analysis. Indeed, Re $\delta_{23}^d$ and Im
$\delta_{23}^d$ play exactly the same role as $\rho$ and $\eta$ in the
SM UT fit: starting from a given (uniform) a priori distribution, the
p.d.f. for Re $\delta_{23}^d$, Im $\delta_{23}^d$ and the observables
we discuss is obtained using the experimental constraints detailed
above.

The gluino-mediated $b \to s$ transitions in the MSSM had already been
investigated by several
authors~\cite{Bertolini:1987pk}--\cite{Causse:2002mu} before the
announcement of $S_{\phi K}$ negative and have been vigorously
reassessed~\cite{Hiller:2002ci}--\cite{Harnik:2002vs} after such
results were announced last Summer. In particular in the works of
refs.~\cite{Khalil:2002fm,Kane:2002sp} the correlation between $B \to
\phi K_s$ and $B_s - \bar B_s$ mixing has been investigated making use
of the mass insertion approximation. In ref.~\cite{Ciuchini:2002uv}
the level of accuracy was improved with the inclusion of NLO QCD
corrections and lattice QCD hadronic matrix elements and also in the
correlation of $b \to s$ related processes with the selection of the
$\Delta B=1$ and $\Delta B=2$ phenomena outlined above. As for the
evaluation of $B \to K \pi$ and $B \to \phi K_s$,
ref.~\cite{Kane:2002sp} adopts the BBNS factorization, but without
discussing the possibly large $\Lambda/M_b$ corrections. This may be
the source of some quantitative difference on the $RR$ contributions
to $BR(B \to \phi K_s)$ between ref.~\cite{Ciuchini:2002uv} and
refs.~\cite{Khalil:2002fm,Kane:2002sp}, as we will detail in next
Section. As for the analysis of ref.~\cite{Harnik:2002vs}, this is
performed in the mass eigenstate basis taking a specific down squark
mass matrix (as suggested in SUSY GUT's where the large neutrino
mixing is linked to a large $\tilde b_R \to \tilde s_R$ mixing).
Comparing the results of refs.~\cite{Ciuchini:2002uv} and
\cite{Harnik:2002vs} in the case of $RR$ dominance we find some
discrepancy in particular in the case of large $(\delta_{23}^d)_{RR}$
(see below).  Once again a potential source of discrepancy in
constraining the $\delta^d_{23}$'s from $A_{CP}(B\to \phi K_s)$ is
represented by the delicate evaluation of the matrix elements of the
chromo-dipole operators.

\begin{figure*}[t]
  \begin{center}
    \begin{tabular}{c c}
      \includegraphics[width=0.4\textwidth]{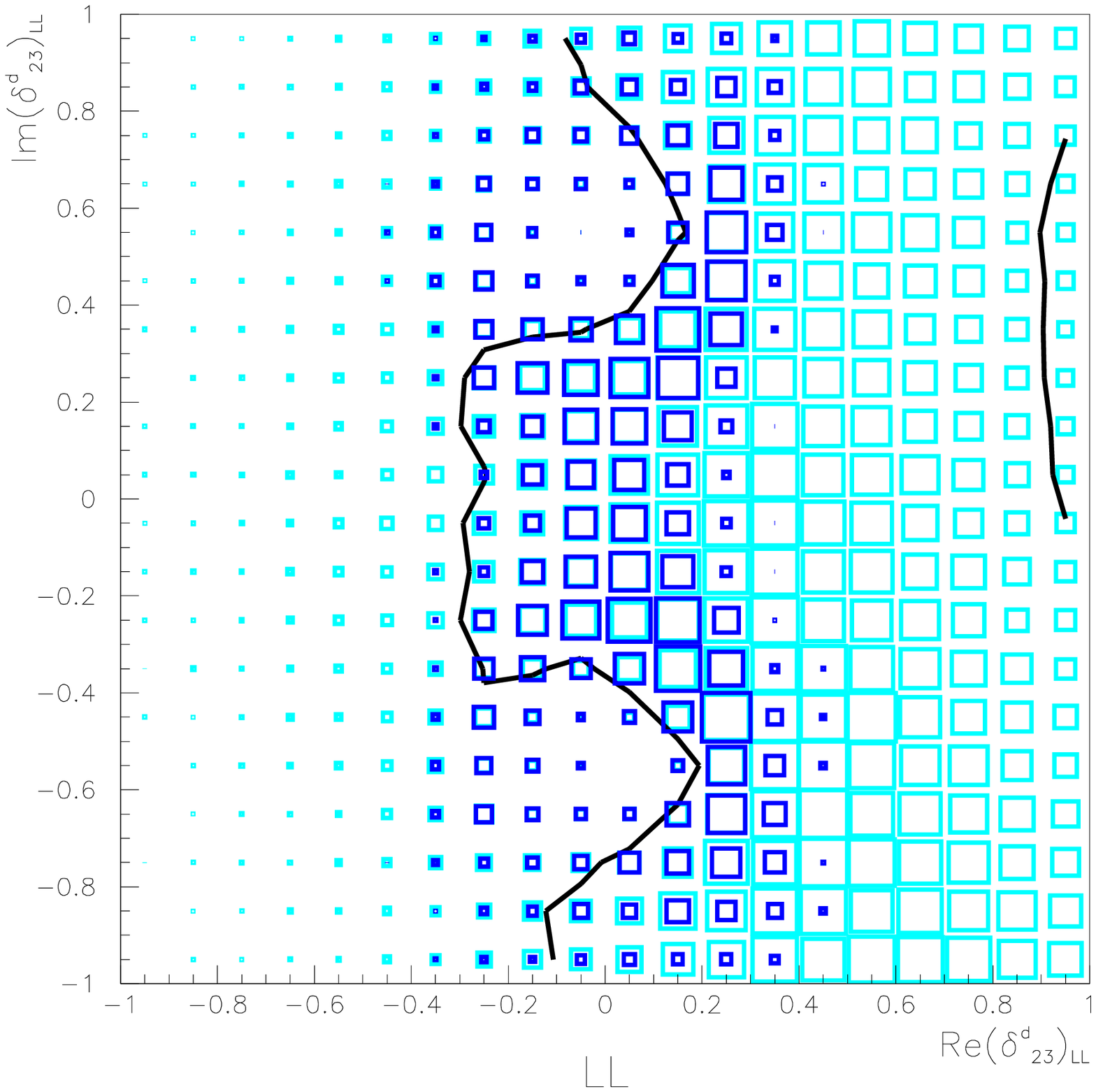} &
      \includegraphics[width=0.4\textwidth]{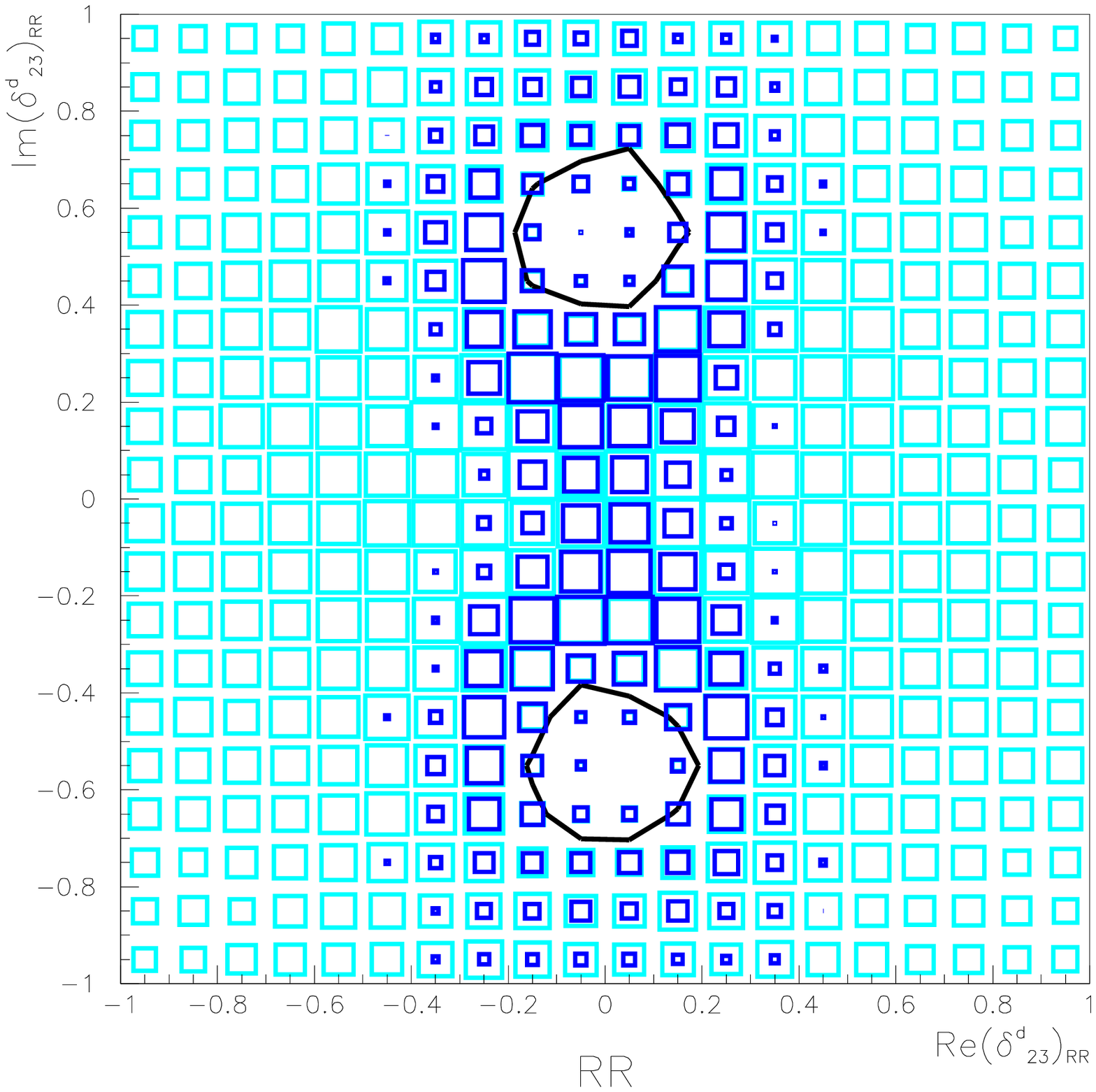} \\
      \includegraphics[width=0.4\textwidth]{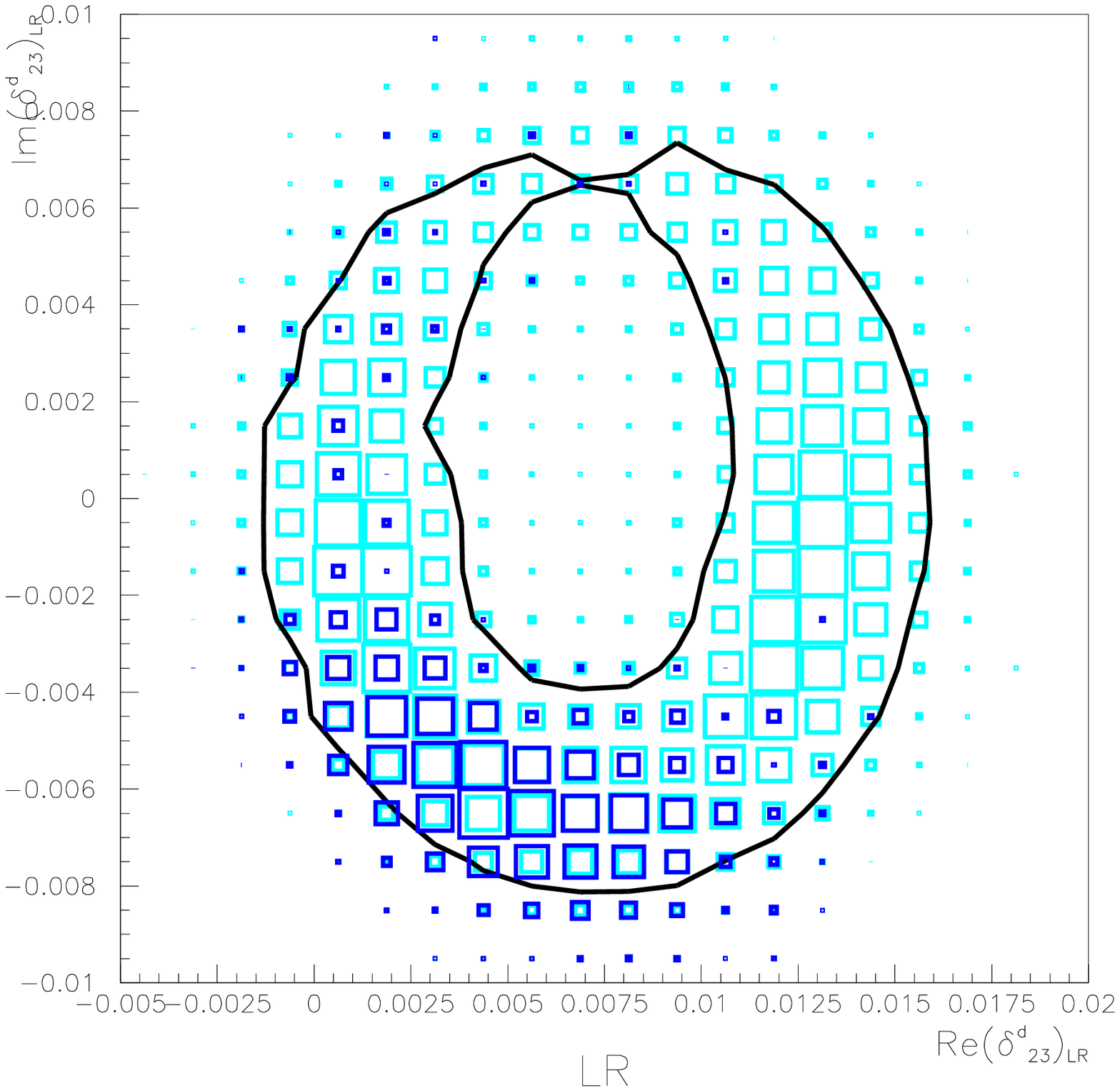} &
      \includegraphics[width=0.4\textwidth]{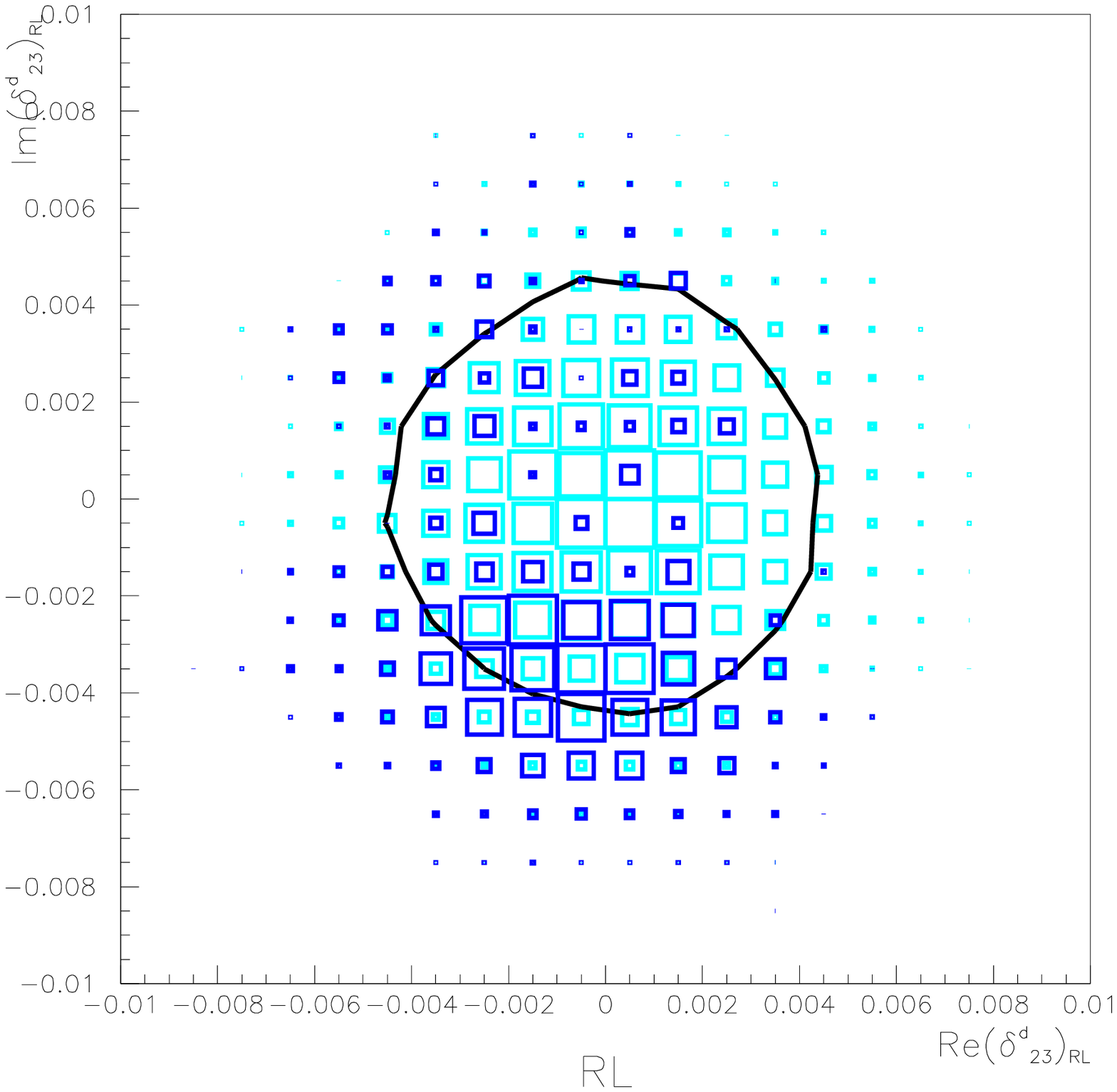} \\ 
    \end{tabular}
  \end{center}
  \caption{Allowed regions in the  
    Re$(\delta^d_{23})_{AB}$--Im$(\delta^d_{23})_{AB}$ space for
    $AB=(LL,RR,LR,RL)$. The black line contains $68 \%$ of the
    weighted events. The darker regions are selected imposing $\Delta
    m_s<20$ ps$^{-1}$ for $LL$ and $RR$ insertions and $S_{\phi K}<0$
    for $LR$ and $RL$ insertions.}
  \label{fig:ranges1}
\end{figure*}

In fig.~\ref{fig:ranges1} we display the p.d.f. in the
Re$(\delta^d_{23})_{AB}$--Im$(\delta^d_{23})_{AB}$ plane in the single
insertion case. Here and in the following plots, larger boxes
correspond to larger numbers of weighted events. Constraints from
$BR(B\to X_s\gamma)$, $A_{CP}(B\to X_s\gamma)$, $BR(B\to X_s \ell^+
\ell^-)$ and the lower bound on $\Delta M_s$ have been applied, as
discussed above. The darker regions are selected imposing the further
constraint $\Delta M_s<20$~ps$^{-1}$ for $LL$ and $RR$ insertions and
$S_{\phi K}<0$ for $LR$ and $RL$ insertions.  For helicity conserving
insertions, the constraints are of order $1$. A significant reduction
of the allowed region appears if the cut on $\Delta M_s$ is imposed.
The asymmetry of the $LL$ and $LR$ plots is due to the interference
with the SM contribution. In the helicity flipping cases, constraints
are of order $10^{-2}$. For these values of the parameters, $\Delta
M_s$ is unaffected. We show the effect of requiring $S_{\phi K}<0$: it
is apparent that a nonvanishing Im$\,\delta_{23}^d$ is needed to meet
this condition.

\begin{figure*}[t]
  \begin{center}
    \begin{tabular}{c c}
      \includegraphics[width=0.4\textwidth]{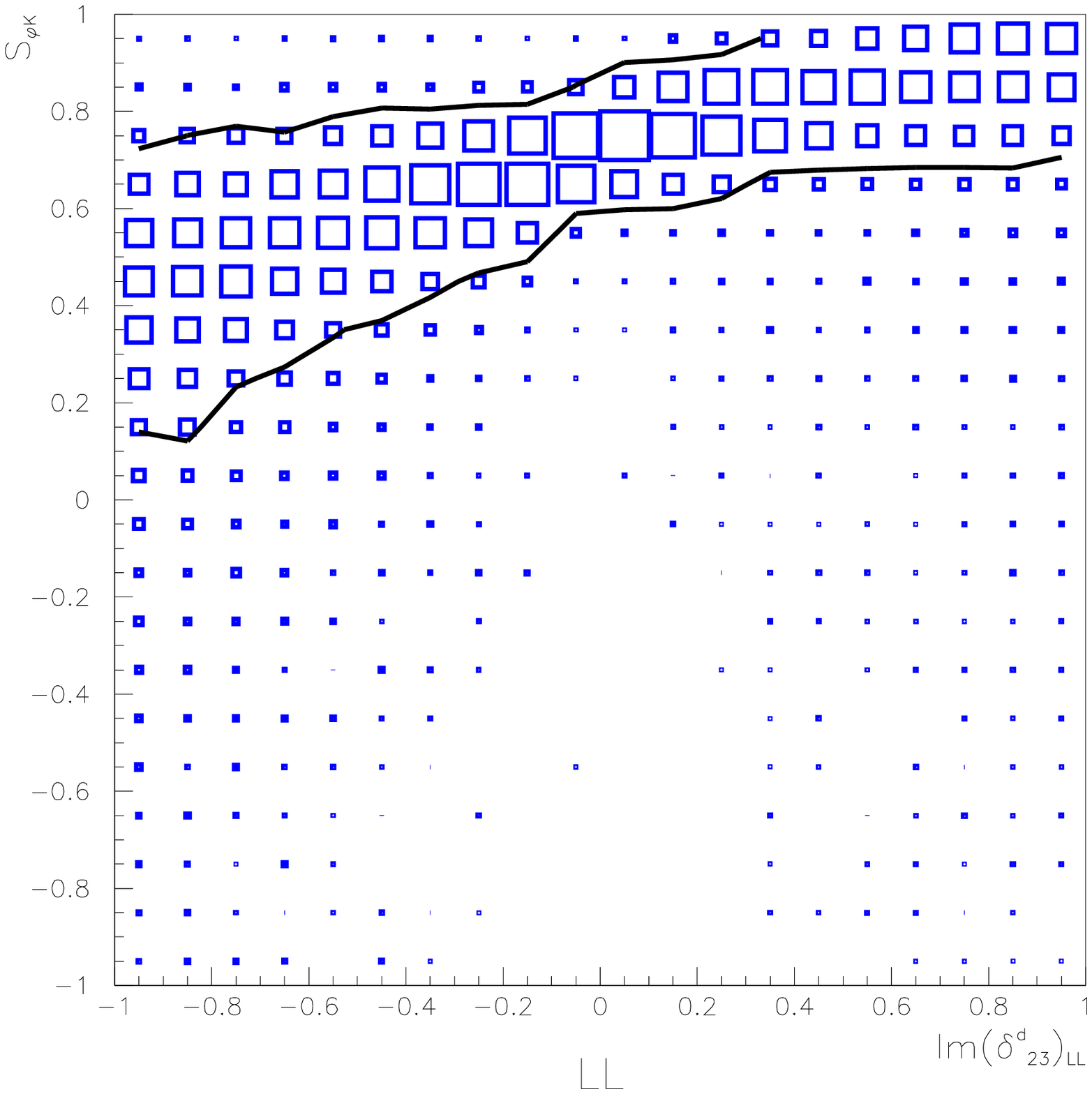} &
      \includegraphics[width=0.4\textwidth]{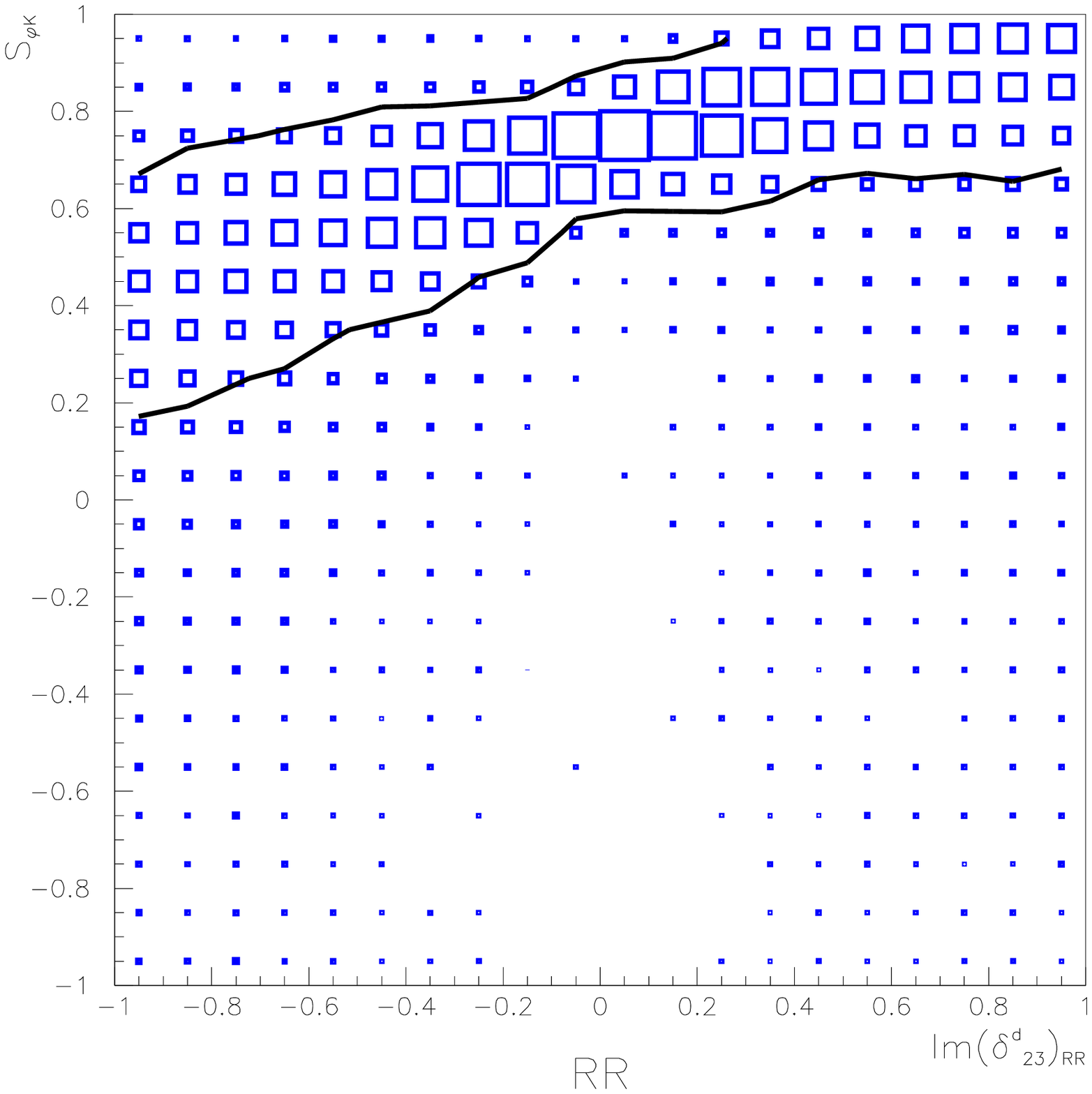} \\
      \includegraphics[width=0.4\textwidth]{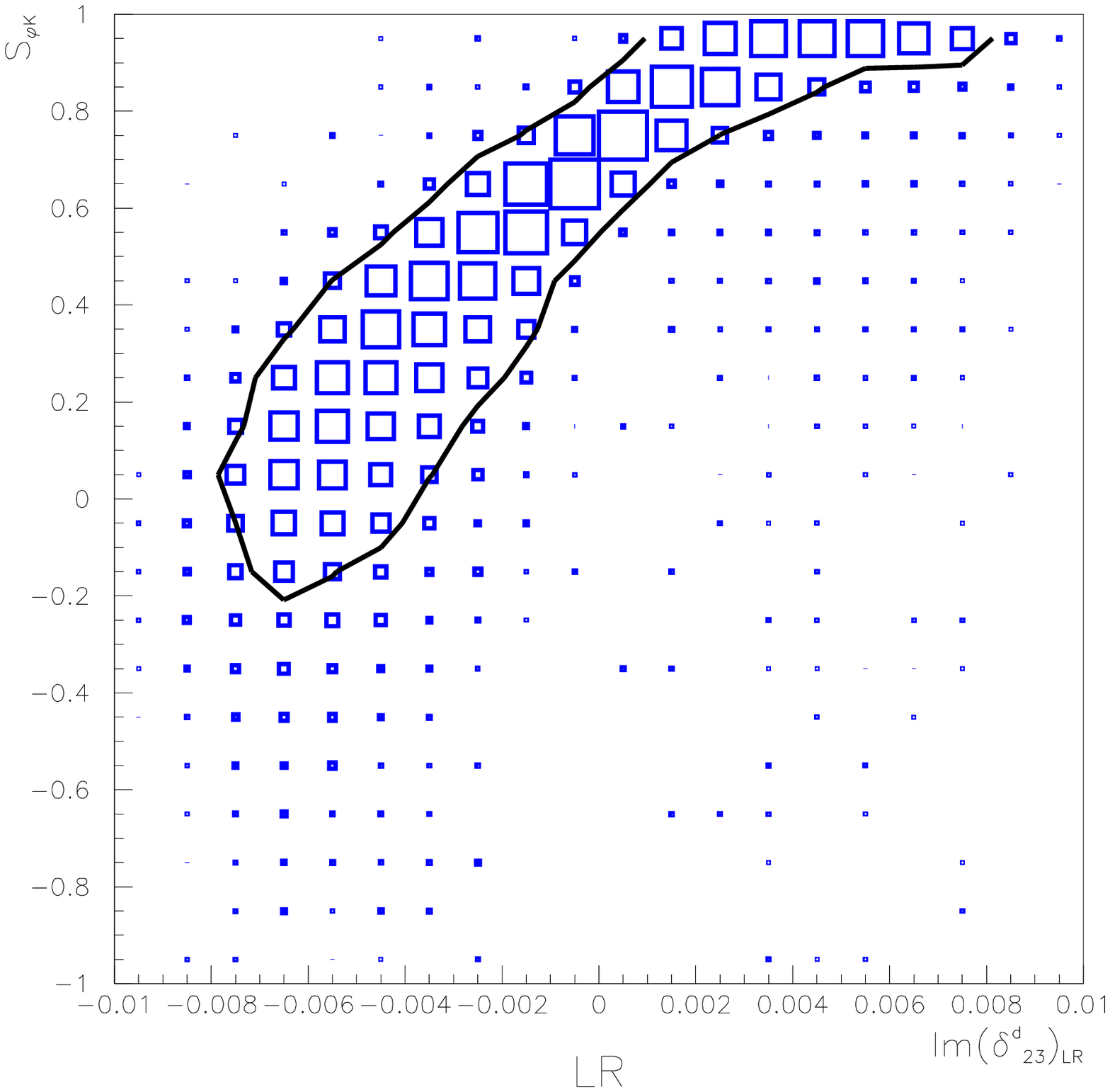} &
      \includegraphics[width=0.4\textwidth]{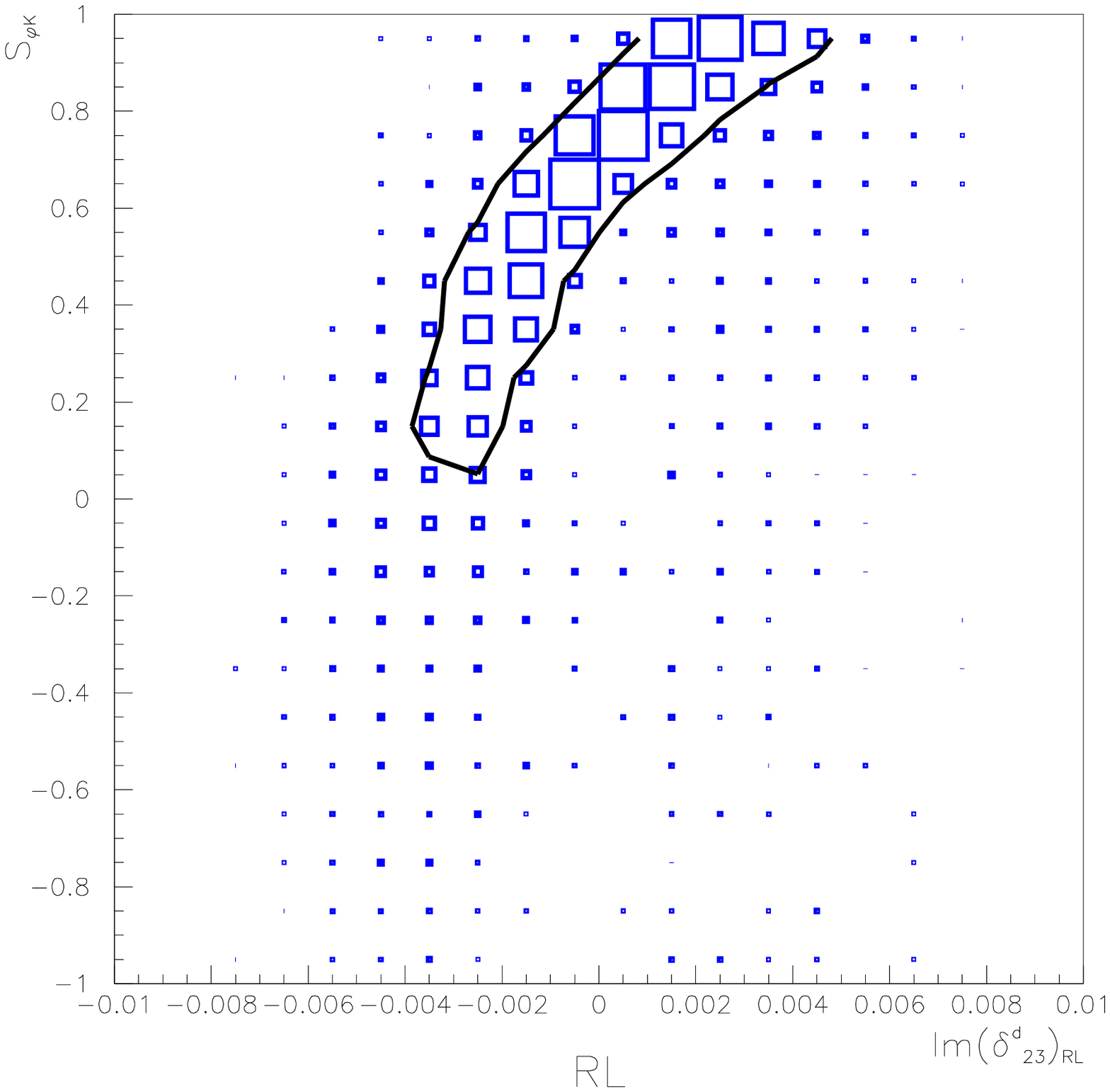} \\ 
    \end{tabular}
  \end{center}
  \caption{Correlations between $S_{\phi K}$ and Im$(\delta^d_{23})_{AB}$ 
    for  $AB=(LL,RR,LR,RL)$. The black line
    contains $68 \%$ of the weighted events.
    }
  \label{fig:sinim1}
\end{figure*}

\begin{figure*}[t]
  \begin{center}
    \begin{tabular}{c c c}
      \includegraphics[width=0.31\textwidth]{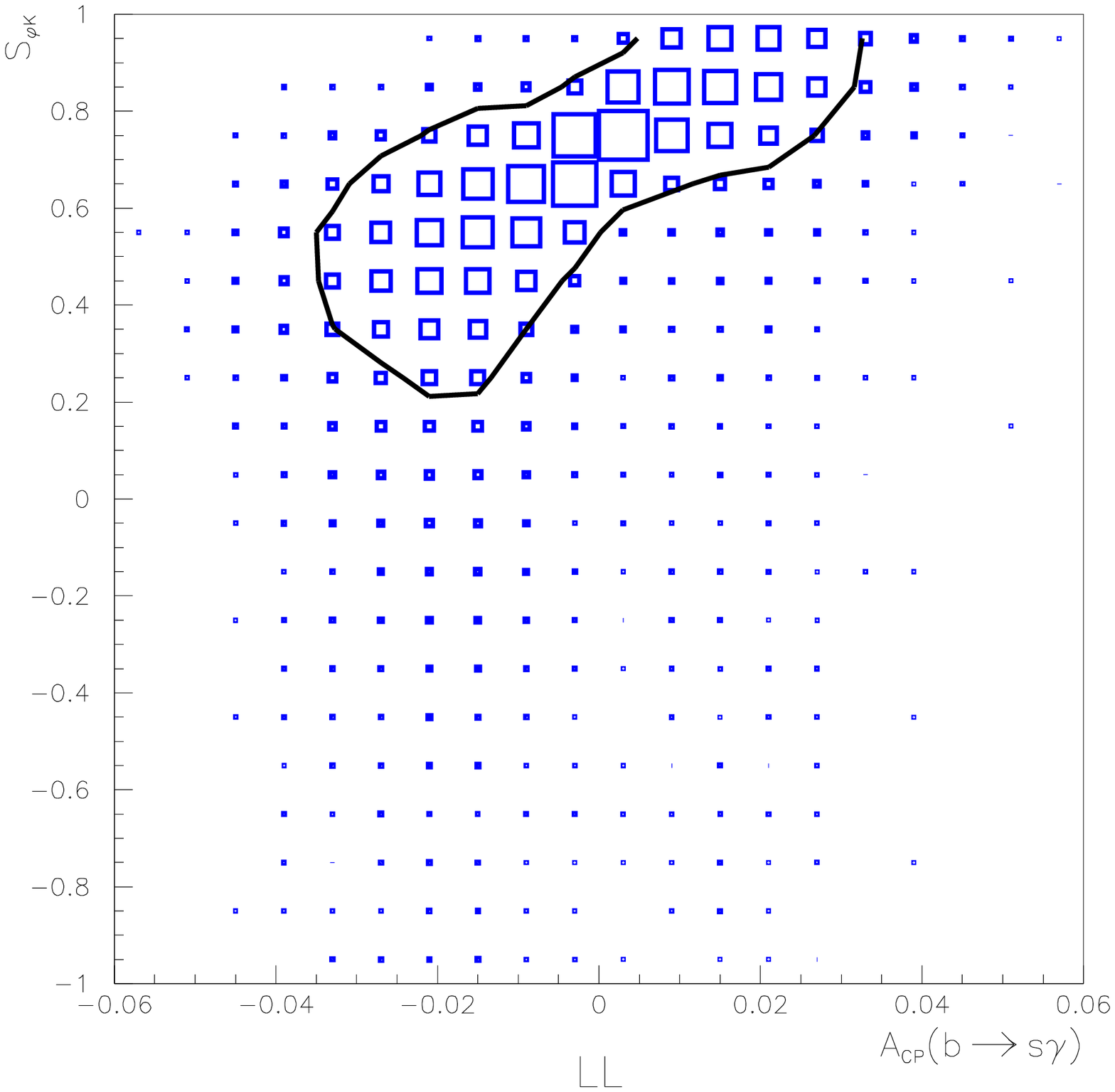} &
      \includegraphics[width=0.31\textwidth]{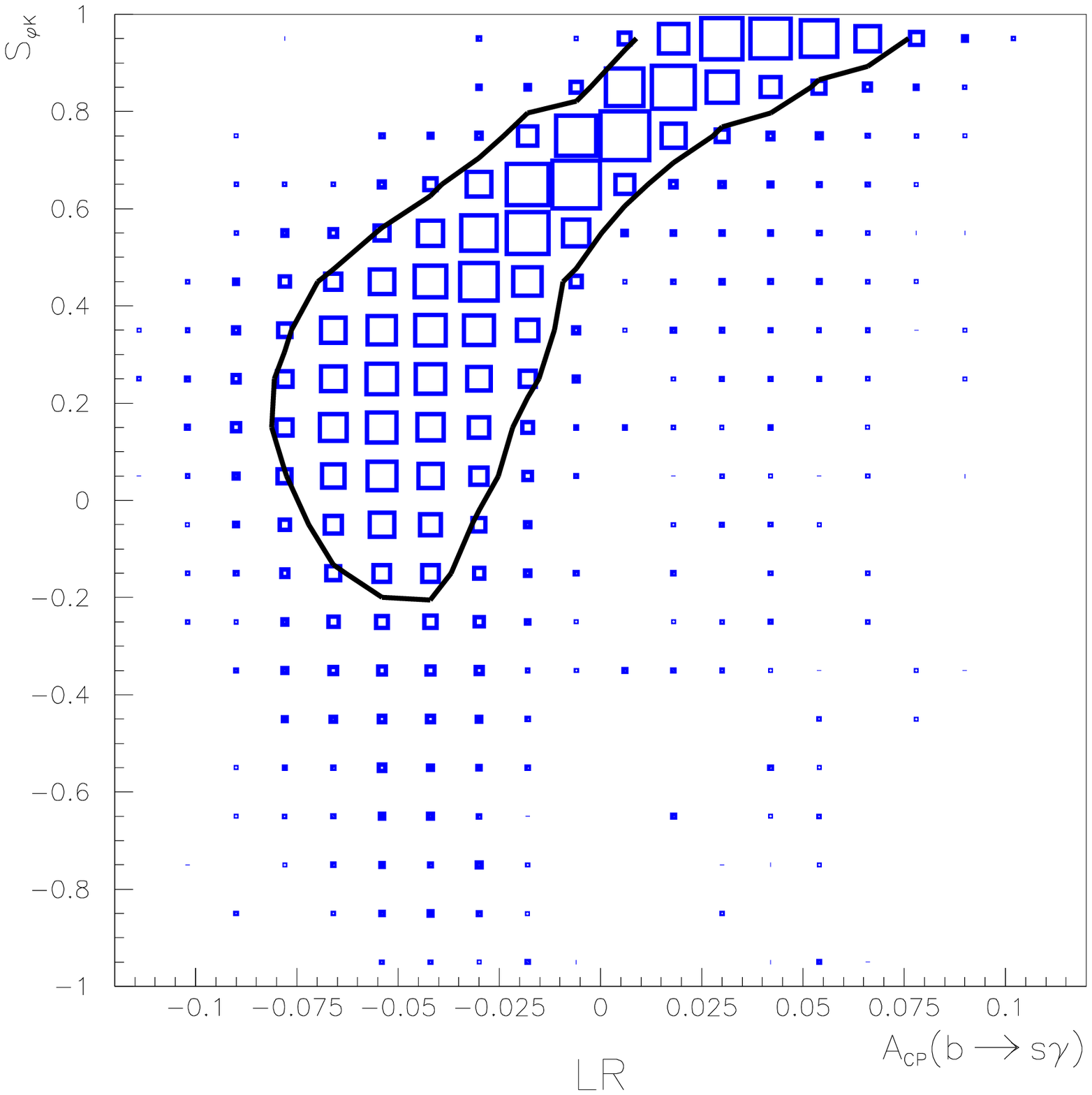} &
      \includegraphics[width=0.31\textwidth]{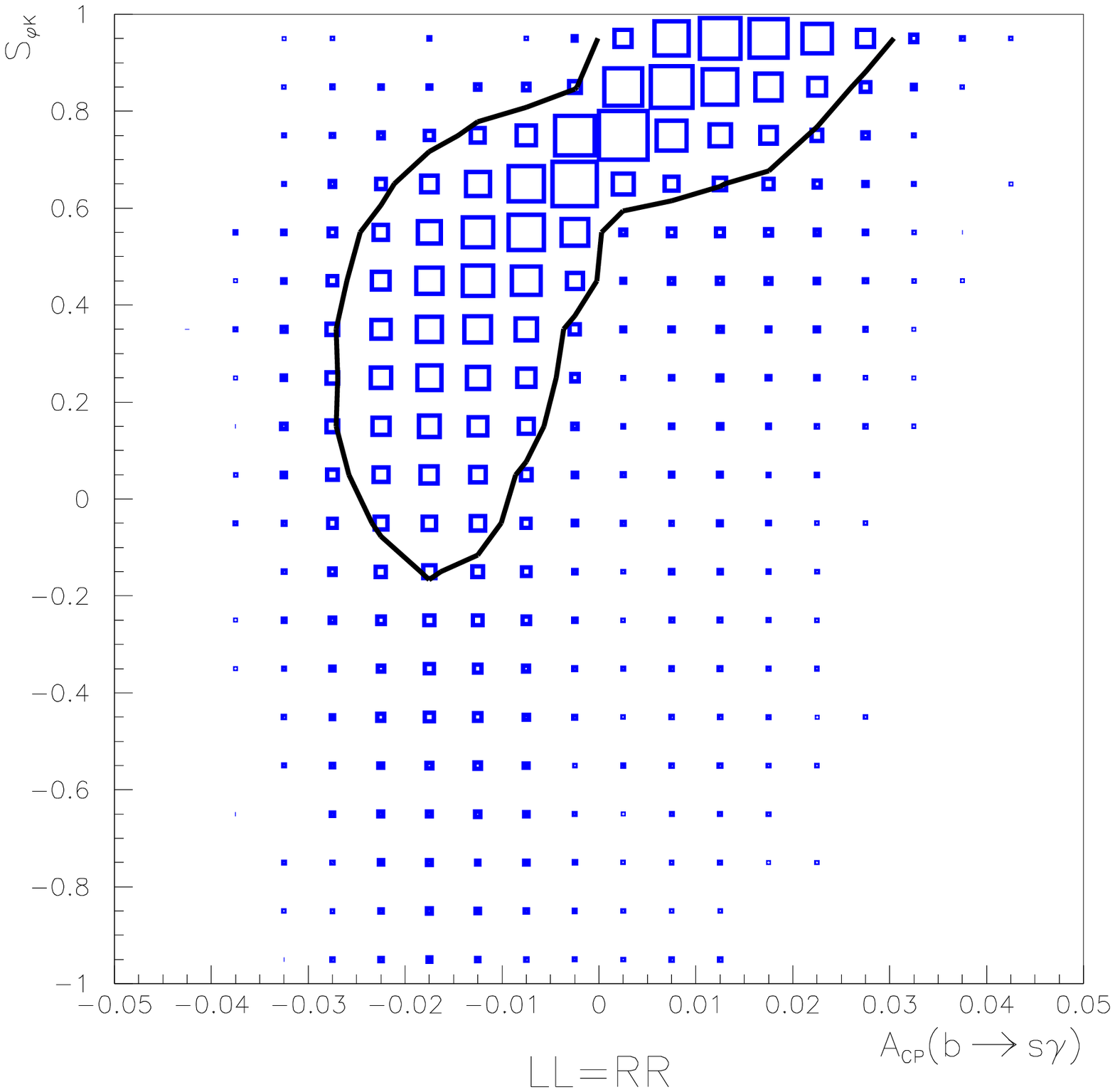}
      \\
    \end{tabular}
  \end{center}
  \caption{Correlation between $S_{\phi K}$ and $A_{CP}(b\to s\gamma)$
    for SUSY mass insertions $(\delta^d_{23})_{AB}$ with
    $AB=(LL,LR,LL=RR)$. The black line
    contains $68 \%$ of the weighted events.}
  \label{fig:sinabsg}
\end{figure*}

\begin{figure*}[t]
  \begin{center}
    \begin{tabular}{c c}
      \includegraphics[width=0.4\textwidth]{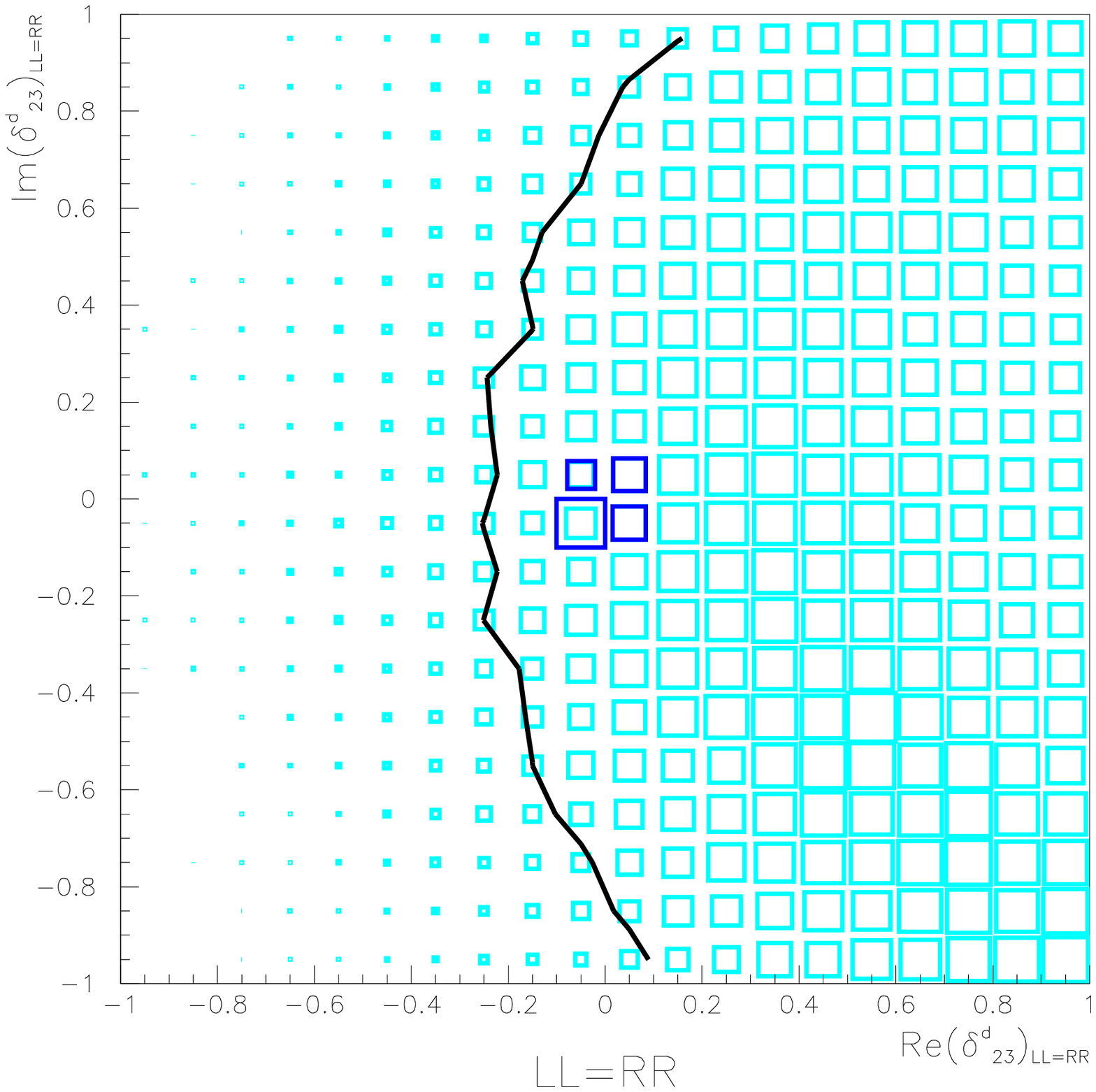} &
      \includegraphics[width=0.4\textwidth]{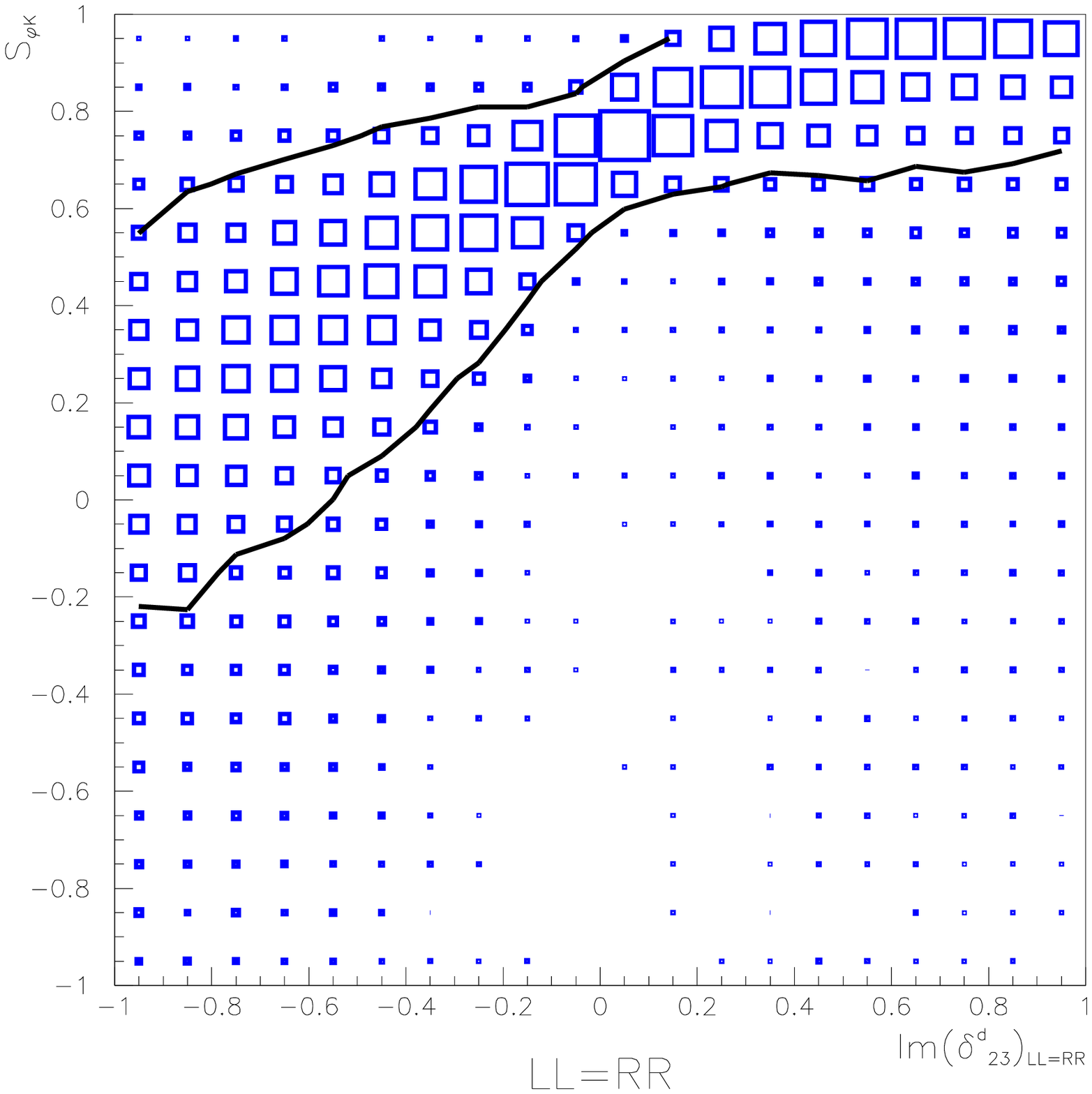}
      \\
    \end{tabular}
  \end{center}
  \caption{Same as in
    figs.~\protect\ref{fig:ranges1}--\protect\ref{fig:sinim1} for the
    double insertion case $LL=RR$. The black line contains $68 \%$ of
    the weighted events. The darker region in the plot on the left is
    selected imposing $\Delta m_s<20$ ps$^{-1}$.}
  \label{fig:double}
\end{figure*}

In figs.~\ref{fig:sinim1}--\ref{fig:sinabsg}, we study the
correlations of $S_{\phi K}$ with Im$(\delta^d_{23})_{AB}$ and
$A_{CP}(B\to X_s\gamma)$ for the various SUSY insertions considered in
the present analysis. The reader should keep in mind that, in all the
results reported in figs.~\ref{fig:sinim1}--\ref{fig:sinabsg}, the
hadronic uncertainties affecting the estimate of $S_{\phi K}$ are not
completely under control. Low values of $S_{\phi K}$ can be more
easily obtained with helicity flipping insertions, in particular in
the $RL$ case. A deviation from the SM value for $S_{\phi K}$ requires
a nonvanishing value of Im$\,(\delta^d_{23})_{AB}$ (see
figs.~\ref{fig:sinim1} and \ref{fig:double}), generating, for those
channels in which the SUSY amplitude can interfere with the SM one, an
$A_{CP}(B\to X_s\gamma)$ at the level of a few percents in the $LL$ and
$LL=RR$ cases, and up to the experimental upper bound in the $LR$ case
(see fig.~\ref{fig:sinabsg}).

Finally, fig.~\ref{fig:double} contains the same plots as
fig.~\ref{fig:ranges1}--\ref{fig:sinim1} in the case of the double
mass insertion $(\delta^d_{23})_{LL}=(\delta^d_{23})_{RR}$.  In this
case, the constraints are still of order $1$, but the contribution to
$\Delta M_s$ is huge, due to the presence of operators with mixed
chiralities. This can be seen from the smallness of the dark region
selected by imposing $\Delta M_s<20$ ps$^{-1}$.

\section{Where to look for SUSY}
A crucial question naturally arises at this point: what are the more
promising processes to reveal some signal of low energy SUSY among the
FCNCs involving $b \to s$ transitions?  For this purpose, it is useful
to classify different ``classes of MSSM'' according to the
``helicities'' $LL$, $RR$, etc, of the different $\delta_{23}^d$'s.

The BaBar and BELLE Collaborations have recently reported the
time-dependent CP asymmetry in $B_d(\bar B_d) \to \phi K_s$.  While
$\sin 2 \beta$ as measured in the $B \to J/\psi K_s$ channel is $0.734
\pm 0.054$ (in agreement with the SM
prediction~\cite{Stocchi:2002yi}), the combined result from both
collaborations for the corresponding $S_{\phi K}$ of $B_d \to \phi
K_s$ is $-0.39\pm 0.41$ \cite{Aubert:2002nx} with a $2.7 \sigma$
discrepancy between the two results. In the SM, they should be the
same up to doubly Cabibbo suppressed terms. Obviously, one should be
very cautious before accepting such result as a genuine indication of
NP.  Nonetheless, the negative value of $S_{\phi K}$ could be due to
large SUSY CP violating contributions. Then, one can wonder which
$\delta$'s are relevant to produce such enhancement and, even more
important, which other significant deviations from the SM could be
detected.

\subsection{\boldmath$RR$ and \boldmath$LL$ cases}
\begin{figure*}[t]
  \begin{center}
    \begin{tabular}{c c}
      \includegraphics[width=0.4\textwidth]{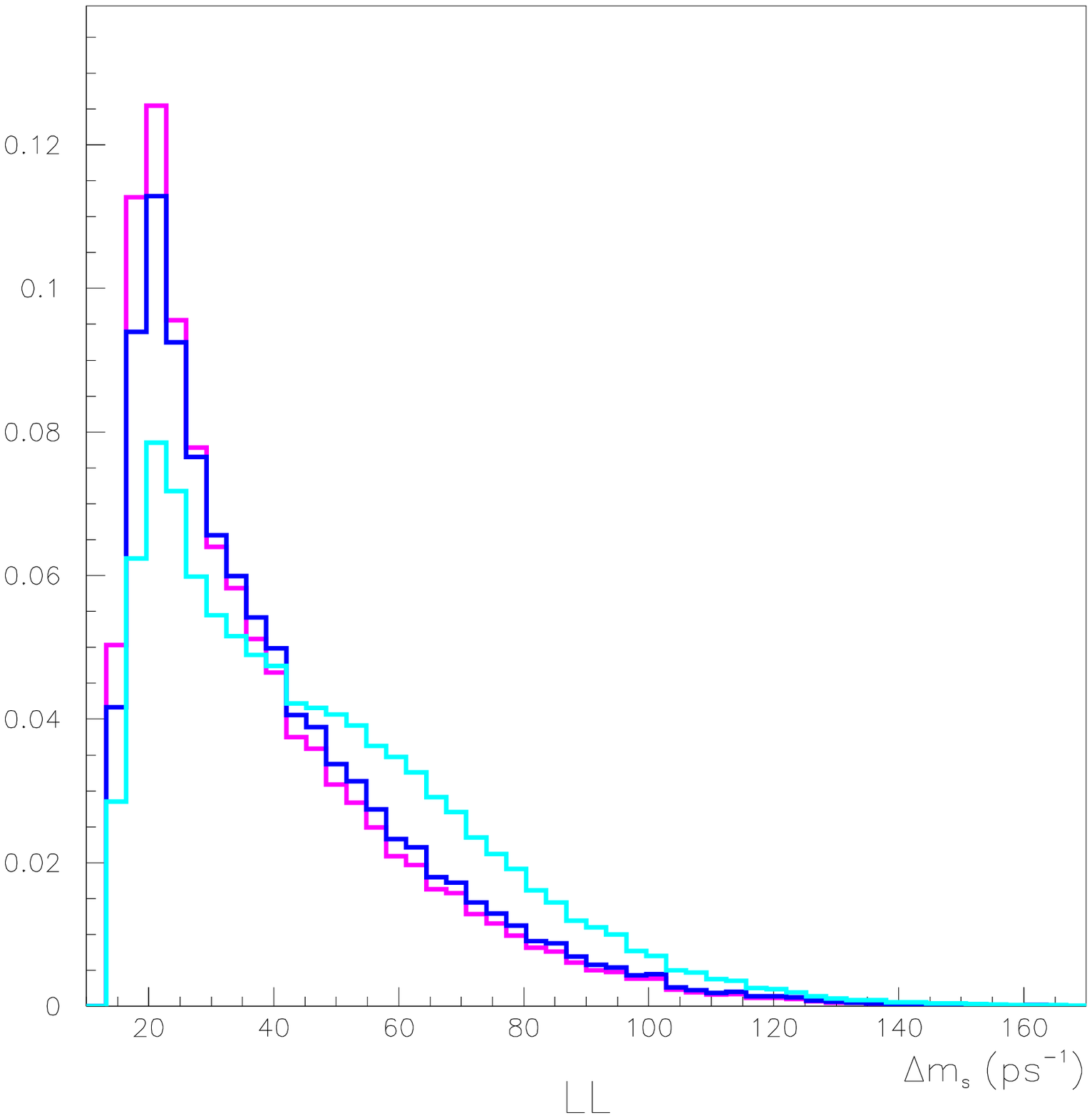} &
      \includegraphics[width=0.4\textwidth]{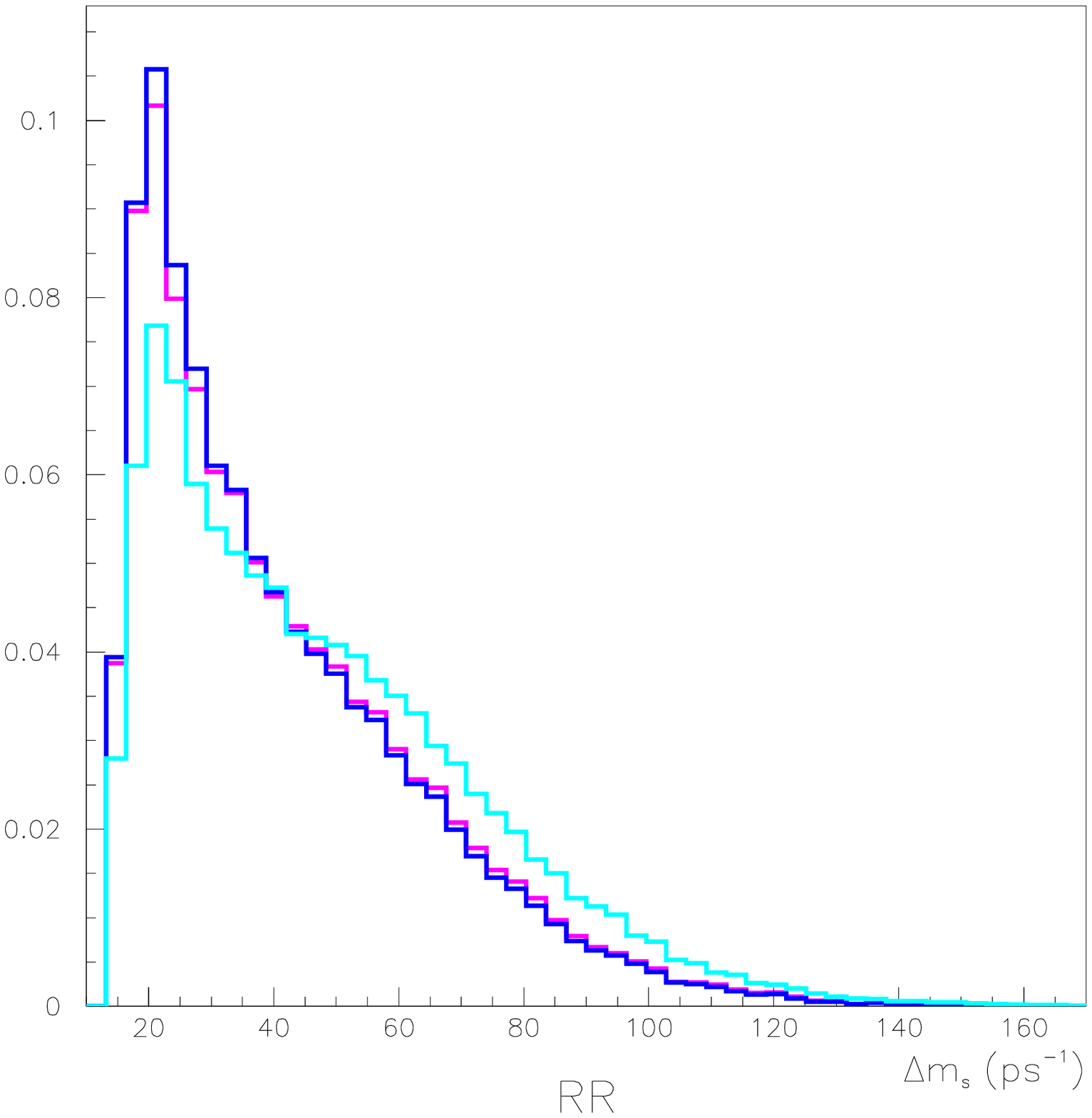} \\
      \includegraphics[width=0.4\textwidth]{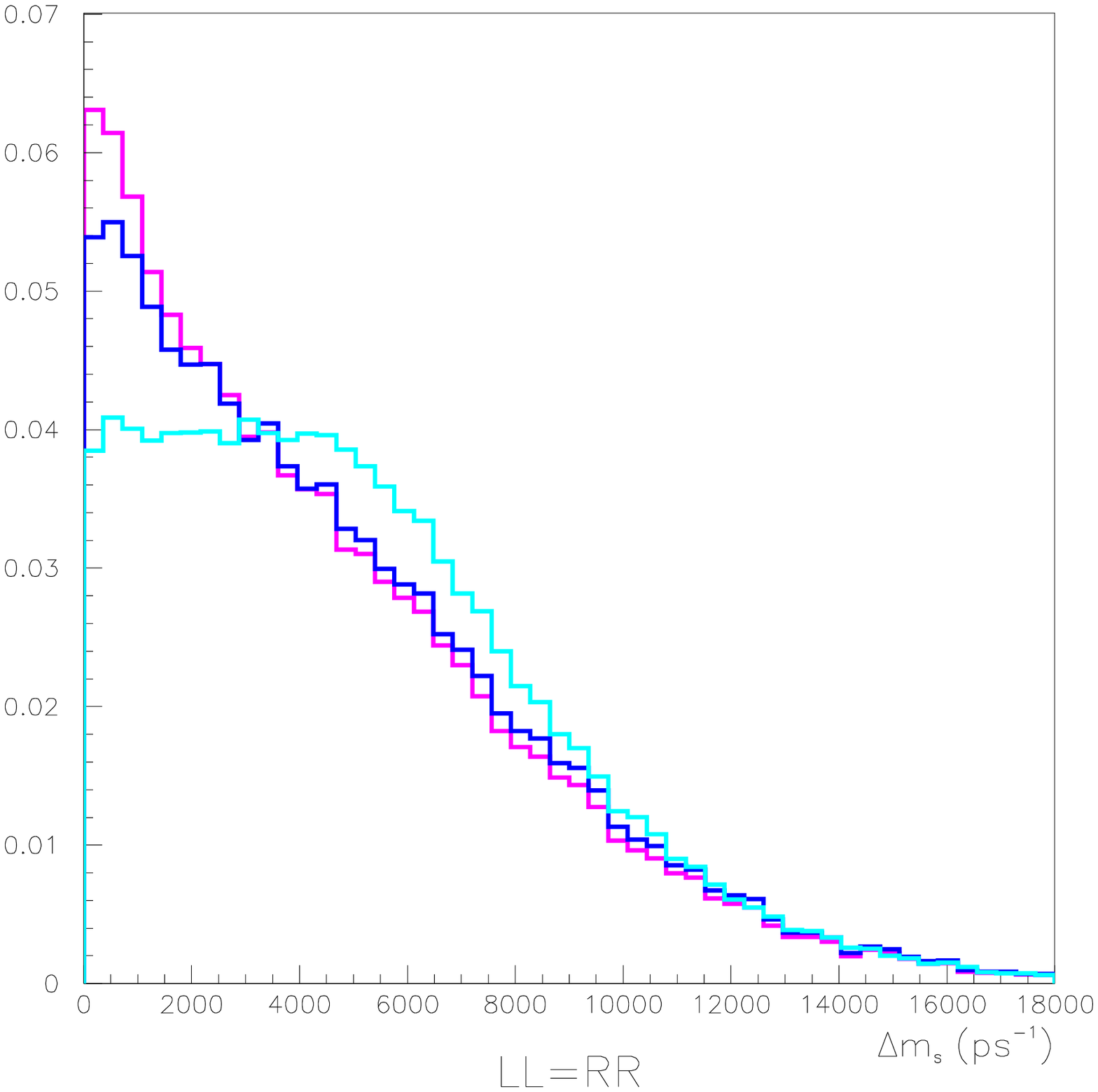} &
      \includegraphics[width=0.4\textwidth]{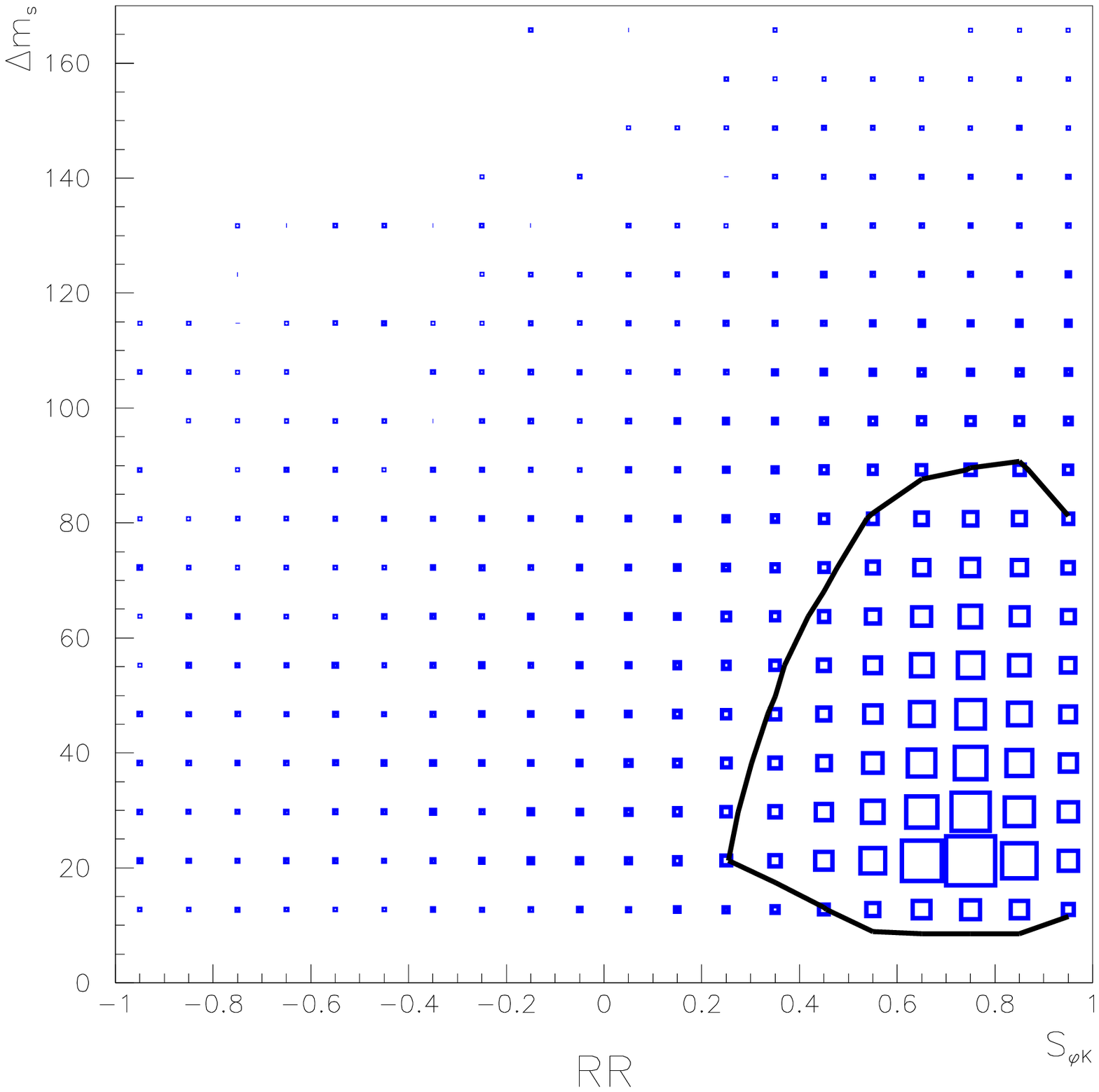} \\
      \\
    \end{tabular}
  \end{center}
  \caption{Distributions of $\Delta M_s$ for SUSY mass 
    insertions $(\delta^d_{23})_{AB}$ with $AB=(LL,RR,LL=RR)$.
    Different curves correspond to the inclusion of constraints from
    $B\to X_s\gamma$ only (magenta), $B\to X_s l^+l^-$ only (cyan) and
    all together (blue).  Lower right: correlation between $\Delta
    M_s$ and $S_{\phi K}$ in the $RR$ case.  }
  \label{fig:dms}
\end{figure*}

We start discussing the $RR$ case. As shown in Fig.~\ref{fig:sinim1}
(upper right), although values of $S_{\phi K}$ in the range predicted
by the SM are largely favoured, still pure $\delta_{RR}$ insertions
are able to give rise to a negative $S_{\phi K}$ in agreement with the
results of BaBar and BELLE quoted above.  On this point we seem to
agree with the conclusions of ref.~\cite{Harnik:2002vs}, while being
in disagreement with refs.~\cite{Kane:2002sp} and
\cite{Khalil:2002fm}. As for the $B_s - \bar B_s$ mixing, the
distribution of $\Delta M_s$ is peaked at the SM value, but it has a
long tail at larger values, up to $\sim 120$ ps$^{-1}$ for our choice of
the range of $\delta_{RR}$. In addition, we find that the expected
correlation requiring large $\Delta M_s$ for negative $S_{\phi K}$ is
totally wiped out by the large uncertainties (see fig.~\ref{fig:dms},
lower right).  In this respect, we are at variance with
ref.~\cite{Harnik:2002vs}, where it was emphasized that if the $RR$
squark mixing yields the large deviation from the SM for the value of
$S_{\phi K}$, then a huge contribution to the $B_s$ mixing should
necessarily follow making such oscillation unobservable at Tevatron.
Hence, in the $RR$ case it is possible to have a strong discrepancy
between $\sin 2\beta$ and $S_{\phi K}$ whilst $B_s-\bar B_s$
oscillations proceed as expected in the SM (thus, being observable in
the Run II of Tevatron).
To conclude the discussion of the $RR$ case, we expect the CP
asymmetry in $B \to X_s \gamma$ to be as small as in the
SM, while, differently from the SM, the
time-dependent CP asymmetry in the decay channel $B_s \to J/\psi \phi$
is expected to be large.

We now move on to discuss the $LL$ insertion.  A major difference with
the previous case concerns the SUSY contributions to $B \to X_s
\gamma$. The $LL$ insertion contributes to the same operator which is
responsible for $B \to X_s \gamma$ in the SM and hence the SM and SUSY
amplitudes interfere. As a consequence, the rate tends to be larger
than the $RR$ case and, moreover, a CP asymmetry can be generated up
to $5 \%$ (see fig.~\ref{fig:sinabsg}, left). However, given the
uncertainties, the correlation of $A_{CP}(B\to X_s\gamma)$ with
$S_{\phi K}$ is not very stringent. As can be seen from the figure,
negative values of $S_{\phi K}$ do not necessarily correspond to
non-vanishing $A_{CP}(B\to X_s\gamma)$, although typical values are
around $2\%$. Also, the constraint coming from the present measurement
of the CP asymmetry is not very effective, as can be seen for instance
from the distribution of $\Delta M_s$ in fig.~\ref{fig:dms} which is
quite similar to the $RR$ case. Finally, one expects also in this case
to observe CP violation in $B_s \to J/\psi \phi$ at hadron colliders.

\subsection{\boldmath$LR$ and \boldmath$RL$ cases}
In these cases negative values of $S_{\phi K}$ can be easily obtained
(although a positive $S_{\phi K}$ is favoured,
cfr.~Fig.~\ref{fig:sinim1}, bottom row). The severe bound on the $LR$
mass insertion imposed by BR$(B \to X_s \gamma)$ (and $A_{CP}(B \to
X_s \gamma)$ in the $LR$ case) prevents any enhancement of the $B_s -
\bar B_s$ mixing as well as any sizeable contribution to $A_{CP}(B_s
\to J/\psi \phi)$.

Notice that the $LR$ mass insertion contributes to $b_R \to s_L
\gamma$, much like the SM. The interference with the SM amplitude
produces the 'hole' in fig.~\ref{fig:ranges1}, lower left. $A_{CP}(B
\to X_s \gamma)$ as large as $5$--$10$ \% is attainable in this case
(Fig.~\ref{fig:sinabsg}, upper right), offering a potentially
interesting hint for NP. On the contrary, the $RL$ mass insertion
contributes to $b_L \to s_R \gamma$ and thus it does not interfere.
Consequently, the CP asymmetry is as small as in the SM.

\subsection{Double mass insertion: $(\delta_{23})_{LL}
  =(\delta_{23})_{RR}$ case}
The main feature of this case is the huge enhancement of $\Delta M_s$
which is made possible by the contribution of the double insertion $LL$
and $RR$ in the box diagrams to operators with mixed chiralities
(Fig.~\ref{fig:dms}, lower left). Differently from all the previous
cases, we are facing a situation here where $A_{CP}(B \to \phi K_s)$ at
its present experimental value should be accounted for by the presence
of SUSY, while $\Delta M_s$ could be so large that the $B_s - \bar
B_s$ mixing could escape detection not only at Tevatron, but even at
BTeV or LHCB. Hence, this would be a case for remarkable signatures of
SUSY in $b \to s$ physics.

Finally, we remark that in the $LR$ and $RL$ cases, since for
$m_{\tilde g}=m_{\tilde q}=350$ GeV the constraints on the
$\delta_{23}^d$'s are of order $10^{-2}$, the same phenomenology in
$\Delta B=1$ processes can be obtained at larger values of mass
insertions and of squark and gluino masses, while contributions to
$\Delta B=2$ processes become more important for larger masses. In the
remaining cases, where the limits on $\delta_{23}^d$ at $m_{\tilde
  g}=m_{\tilde q}=350$ GeV are of order $1$, the SUSY effects clearly
weaken when going to higher values of sparticle masses.

Complementary information on SUSY contributions to $b \to s$
transitions can be found in
refs.~\cite{Baek:2003kb,Chakraverty:2003uv}, where chargino
contributions to $A_{CP}(B \to \phi K_s)$ have been considered, and in
ref.~\cite{Khalil:2003bi}, where the correlation between SUSY
contributions to $A_{CP}(B \to \phi K_s)$ and $A_{CP}(B \to
\eta^\prime K_s)$ have been discussed.

\section{Outlook}
 
Our results confirm that FCNC and CP violation in
physics involving $b \to s$ transitions still offer
opportunities to disentangle effects genuinely due to NP. In
particular the discrepancy between the amounts of CP violation in the
two $B_d$ decay channels $J/\psi K_s$ and $\phi K_s$ can be accounted
for in the MSSM while respecting all the existing constraints in $B$
physics, first of all the $BR(B \to X_s \gamma)$. The relevant question
is then which processes offer the best chances to provide other hints
of the presence of low-energy SUSY.

First, it is mandatory to further assess the time-dependent CP
asymmetry in the decay channel $B \to \phi K_s$.  If the measurement
will be confirmed, then this process would become decisive in
discriminating among different MSSM realizations. Although, as we have
seen, it is possible to reproduce the negative $S_{\phi K}$ in a
variety of different options for the SUSY soft breaking down squark
masses, the allowed regions in the SUSY parameter space are more or
less tightly constrained according to the kind of $\delta_{23}^d$ mass
insertion which dominates.

In order of importance, it then comes the measurement of the $B_s -
\bar B_s$ mixing. Finding $\Delta M_s$ larger than $20$ ps$^{-1}$
would hint at NP. $RR$ or $LL$ could account for a $\Delta M_s$ up to
$\sim 120$ ps$^{-1}$. Larger values would call for the double insertion
$LL=RR$ to ensure such a huge enhancement of $\Delta M_s$ while
respecting the constraint on $BR(B \to X_s \gamma)$. An interesting
alternative would arise if $\Delta M_s$ is found as expected in the SM
while, at the same time, $S_{\phi K}$ is confirmed to be negative.
This scenario would favour the $RL$ or $LR$ possibility, even though
all other cases but $LL=RR$ do not necessarily lead to large $\Delta
M_s$.

Keeping to $B_d$ physics, we point out that the CP asymmetry in $B \to
X_s \gamma$ remains of utmost interest.  This asymmetry is so small in
the SM that it should not be possible to detect it. We have seen that
in particular with $LR$ insertions such asymmetry can be enhanced up
to $10$ \% making it possibly detectable in a not too distant future.

Finally, once we will have at disposal large amounts of $B_s$, it will
be of great interest to study processes which are mostly CP conserving
in the SM, while possibly receiving large contributions from SUSY. In
the SM the amplitude for $B_s - \bar B_s$ mixing does not have an
imaginary part up to doubly Cabibbo suppressed terms and decays like
$B_s \to J/\psi \phi$ also have a negligible amount of CP violation.
Quite on the contrary, if the measured negative $S_{\phi K}$ is due to
a large, complex $\delta_{23}^d$ mass insertion, we expect some of the
above processes to exhibit a significant amount of CP violation. In
particular, in the case of $RR$ insertions, both the $b \to s$
amplitudes and the $B_s$ mixing would receive non negligible
contributions from Im$\,\delta_{23}^d$, while, if the $LR$ or $RL$
insertions are dominant, we do not expect any sizable contribution to
$B_s$ mixing. Still, the SUSY contribution to CP violation in the $B_s
\to J/\psi \phi$ decay amplitude could be fairly large.

\end{document}